\documentclass[submission,copyright,creativecommons]{eptcs}
\providecommand{\event}{VPT 2020} 
\usepackage{breakurl}             
\usepackage{underscore}           
\usepackage{graphicx} 
\usepackage{subfigure}
\usepackage{epsf}
\usepackage{proof}
\usepackage{amssymb}

\newcommand{\system}[1]{{\sc #1}}
\newcommand{\prolog}[1]{{\tt #1}}

\newcommand{\prob}{\textsc{ProB}}

\newcommand{\ignore}[1]{}

\usepackage{listings}
\usepackage{xcolor}

\colorlet{punct}{red!60!black}
\definecolor{background}{HTML}{EEEEEE}
\definecolor{delim}{RGB}{20,105,176}
\colorlet{numb}{magenta!60!black}

\lstdefinelanguage{json}{
    basicstyle=\normalfont\ttfamily,
    numbers=left,
    numberstyle=\scriptsize,
    stepnumber=1,
    numbersep=8pt,
    showstringspaces=false,
    breaklines=true,
    frame=lines,
    backgroundcolor=\color{background},
    string=[s]{"}{"},
    comment=[l]{:\ "},
    morecomment=[l]{:"},
    literate=
     *{0}{{{\color{numb}0}}}{1}
      {1}{{{\color{numb}1}}}{1}
      {2}{{{\color{numb}2}}}{1}
      {3}{{{\color{numb}3}}}{1}
      {4}{{{\color{numb}4}}}{1}
      {5}{{{\color{numb}5}}}{1}
      {6}{{{\color{numb}6}}}{1}
      {7}{{{\color{numb}7}}}{1}
      {8}{{{\color{numb}8}}}{1}
      {9}{{{\color{numb}9}}}{1}
      {:}{{{\color{punct}{:}}}}{1}
      {,}{{{\color{punct}{,}}}}{1}
      {\{}{{{\color{delim}{\{}}}}{1}
      {\}}{{{\color{delim}{\}}}}}{1}
      {[}{{{\color{delim}{[}}}}{1}
      {]}{{{\color{delim}{]}}}}{1},
}

\lstdefinelanguage{svg}{
    basicstyle=\normalfont\ttfamily,
    numbers=left,
    numberstyle=\scriptsize,
    stepnumber=1,
    numbersep=8pt,
    showstringspaces=false,
    breaklines=true,
    frame=lines,
    backgroundcolor=\color{background},
    string=[s]{"}{"},
}

\lstdefinelanguage{b}{
basicstyle=\normalfont\ttfamily,
    numbers=left,
    numberstyle=\scriptsize,
    stepnumber=1,
    numbersep=8pt,
    showstringspaces=false,
    breaklines=true,
    frame=lines,
    backgroundcolor=\color{background},
    keywordstyle=\color{numb},
    string=[s]{"}{"},
keywords={
    MACHINE,
    DEFINITIONS,
    CONSTANTS,
    PROPERTIES,
    VARIABLES,
    INVARIANT,
    INITIALISATION,
    INCLUDES,
    OPERATIONS,
    SETS,
    NATURAL,
    INTEGER,
    ANY,
    WHERE,
    SELECT,
    BEGIN,
    END,
    THEN, 
    PRE,
    IF,
    ELSE,
    LET,
  },
}

\title{Prolog for Verification, Analysis and Transformation Tools}

\author{Michael Leuschel
\institute{Lehrstuhl Softwaretechnik und Programmiersprachen\\
Institut f\"{u}r Informatik \\
Heinrich-Heine-Universit\"{a}t D\"{u}sseldorf\\
 Universit\"{a}tsstr. 1\\
  D-40225 D\"{u}sseldorf, Germany}
\email{leushel@hhu.de}
}
\def\titlerunning{Prolog for Verification, Analysis and Transformation}
\def\authorrunning{Michael Leuschel}
\begin{document}
\maketitle

\begin{abstract}
This article examines the use of the Prolog language for writing verification, analysis and
 transformation tools.
Guided by experience in teaching and the development of verification tools like \prob{} or specialisation tools
 like {\sc ecce} and {\sc logen}, the article presents an assessment of various aspects of Prolog
  and provides guidelines for using them.
The article shows the usefulness of a few key Prolog features.
In particular, it discusses how to deal with negation at the level of the object programs being verified or analysed.
\end{abstract}

\section{Tools}

Over the years I have written a variety of tools for verification and transformation,
 mainly using the Prolog programming language.
Indeed, over the years, I found out that Prolog is both a convenient language to express
 the semantics of various programming and specification languages as well as
  transformation and verification rules and algorithms.
 
My first intense engagement with Prolog was initiated in my Master's thesis at the KU Leuven, with the goal of writing
a partial evaluator for Prolog.
Initially I was actually inclined to write the partial evaluator in a functional programming language.
Indeed, my initial contacts with Prolog in the AI course at the University of Brussels were not all
that compelling to me: Prolog seemed like a theoretically appealing language, but practically leading
to programs that often either loop or say ``no''.
While I have obviously revised my opinion since then, I encounter this initial
 disappointment and confusion every year in the eyes of some of the
 students attending a logic programming course.
I have also encountered several students who in their bachelor's or master's thesis
 wanted to implement program analysis techniques for Prolog, but were also afraid
 to write the tools and algorithms themselves in Prolog.
I can understand their anxiety, but try to convince them (not always with success), to 
 use Prolog to write their Prolog analysis algorithms.
Indeed, once the initial hurdles are overcome, Prolog is a very nice language for
 program verification and analysis, both in research and teaching.
For example, in my experience,
  the manipulation and transformation of abstract syntax trees
 can often be done much more compactly, reliably and also efficiently (both memory and time wise) in Prolog
 than in more mainstream languages such as Java.
I don't believe that I would have been able to develop and maintain the core of the \prob\ validation tool
\cite{LeuschelButler:FME03,DBLP:journals/sttt/LeuschelB08}
in an imperative or object-oriented language (without re-inventing a Prolog-inspired library).
In the rest of this article I will mention some noteworthy aspects of Prolog for
verification and program analysis tools.
This is a follow-on from the article \cite{Le08PPDP} from 2008, also providing arguments for using declarative
 programming languages for verification.
I will look at three issues in more detail in Section~\ref{sec-interpreters}:
\begin{itemize}
 \item non-determinism, in particular for deterministic languages
 \item unification
 \item how to handle negation
\end{itemize}
I will also re-examine some statements from \cite{Le08PPDP} in Section~\ref{sec-assessment}.


\section{Writing Interpreters and Validation Tools in Prolog}
\label{sec-interpreters}

It is particularly easy to write an interpreter for Prolog in Prolog; the plain vanilla interpreter
consists just of three small clauses  (see, e.g., \cite{HillGallagher:metachapter,AptTurini:MetaBook}).
Prolog is also a convenient language to express the semantics of
 other programming or specification languages.
In my lectures, I encode the operational semantics of a wide range of imperative
 languages in Prolog, ranging from three-address code \cite{Aho:Dragon2} and Java byte code to more complex
 languages.
In research, I found Prolog useful for the semantics of Petri nets \cite{LeuschelLehmann:CL2000,FarwerLeuschel:PPDP04},
 the CSP process algebra in Prolog \cite{LeuschelFontaine:ICFEM08}
 or the B specification language \cite{LeuschelButler:FME03,DBLP:journals/sttt/LeuschelB08}.
A lot of other researchers have encoded various languages in Prolog:
 Verilog \cite{Bowen:TR99}, Erlang \cite{DBLP:conf/hopl/Armstrong07},
 Java Bytecode \cite{DBLP:journals/entcs/Gomez-ZamalloaAP07,DBLP:conf/esop/AlbertAGPZ07,AlbertGG07,DBLP:conf/padl/AlbertGHP07},
 process algebras \cite{XMC:CAV2000}, 
 to name just a few.
More recently, constrained Horn clause programs have become very popular to encode
 imperative programs \cite{DBLP:conf/pldi/GrebenshchikovLPR12} and have led to new techniques such as \cite{DBLP:journals/tplp/KafleGG18}.

\subsection{Non-Determinism}

\newcommand{\interpret}[3]{#1 \stackrel{#2}{\leadsto} #3} 

Prolog's non-determinism is of course very convenient when modelling non-deterministic specification languages.
Operational semantic rules can often be translated to Prolog clauses.
Take, for example, these two inference rules for a prefix operator $\rightarrow$ and an interleaving operator $\|$  in a process algebra inspired by CSP,
 where $\interpret{X}{a}{X'}$ means that the process $X$ can execute the action $a$ and then behave like the process $X'$:

$$\infer[~]{ \interpret{a\rightarrow Y}{a}{Y}  }{ ~ } ~~~~~ \infer[~]{ \interpret{X \| Y}{a}{X' \| Y}  }{ \interpret{X}{a}{X'} } ~~~~~ \infer[~]{ \interpret{X \| Y}{a}{X \| Y'}  }{ \interpret{Y}{a}{Y'} }$$

These rules can be encoded in Prolog as follows, where {\tt trans/3} encodes the ternary semantics relation $\leadsto$:

\begin{footnotesize}
\begin{verbatim}
  trans('->'(A,Y),A,Y).
  trans('||'(X,Y),A,'||'(X2,Y)) :- trans(X,A,X2).
  trans('||'(X,Y),A,'||'(X,Y2)) :- trans(Y,A,Y2).
\end{verbatim}
\end{footnotesize}

We can then determine that the process $a \rightarrow stop \| b\rightarrow stop$ can perform two possible actions:

\begin{footnotesize}
\begin{verbatim}
| ?- trans('||'('->'(a,stop),'->'(b,stop)),A,R).
A = a,
R = '||'(stop,(b->stop)) ? ;
A = b,
R = '||'((a->stop),stop) ? ;
no
\end{verbatim}
\end{footnotesize}

We can also compute the two possible traces of length two:
\begin{footnotesize}
\begin{verbatim}
| ?- trans('||'('->'(a,stop),'->'(b,stop)),A1,_R1), trans(_R1,A2,R2).
A1 = a,
A2 = b,
R2 = '||'(stop,stop) ? ;
A1 = b,
A2 = a,
R2 = '||'(stop,stop) ? 
yes
\end{verbatim}
\end{footnotesize}

It is quite straightforward to perform exhaustive model checking for such specifications in Prolog, in particular
 if we have access to tabling (aka memoization) to detect repeated reachable states (or processes),
 see, e.g., \cite{XMC:CAV2000,LeuschelMassart:LOPSTR99}.

However, Prolog's non-determinism also comes in handy for deterministic imperative languages, when
 moving from interpretation to analysis.
Here is an excerpt of an interpreter for a subset of the Java Bytecode, which I use in my lectures.
Every instruction in the Java Bytecode consists of an opcode (one byte), followed by its arguments.
Java Bytecode uses an operand stack to store arguments to operators and to push results of operators.
For example, the {\tt imul} opcode removes the topmost (integer) values from the stack an pushes
the result of the multiplication back onto the stack.%
\footnote{Java Bytecode is thus also zero-address-code, as such instructions do not take arguments: they implicitly know where the
 operands are and where the result should stored.}

The code fragment below shows the code for the {\tt iconst} opcode to push a constant onto the operator stack,
 {\tt iop} to perform a binary arithmetic operation on the two topmost stack elements,
 {\tt dup} to duplicate the topmost value on the stack, {\tt return} to stop a method (and return void),
 and a conditional {\tt if1} which jumps to a given label if an operator (applied to the topmost stack element and a provided constant) returns true.
 
Every instruction in the Java Bytecdoe consists of an opcode (one byte), followed by its arguments.
As mentioned above, {\tt imul} and {\tt iadd} take no arguments.
However, these opcodes are converted for our interpreter into the generic {\tt iop} instruction
 (which obviously does not exist as such in the Java virtual machine) with the operator as argument.
A similar grouping of opcodes has been performed for the conditional instruction, e.g., the bytecode instruction {\tt ifle 25} gets translated into
 the Prolog term {\tt if1(<=,0,25)} for our interpreter.
Similarly, opcodes like {\tt iconst\_2} take no arguments, but in the Prolog representation below this is represented
 for simplicity as the Prolog term {\tt iconst(2)}.
The Java Bytecode object program is represented by {\tt instr(PC,Opcode,Size)} facts,
 where {\tt PC} is the position (in bytes) of the opcode, 
 {\tt Opcode} the Prolog term describing the opcode, and {\tt Size} the size in bytes of the opcode
 (which is needed to determine the position in bytes of the next opcode).
This is a small artificial program, which computes 2*2 and then decrements the value until it reaches 0:

\begin{footnotesize}
\begin{verbatim}
    instr(0,iconst(2),1).
    instr(1,iconst(2),1).
    instr(2,iop(*),1).
    instr(3,iconst(-1),1).
    instr(4,iop(+),1).
    instr(5,dup,1).
    instr(6,if1('>',0,3),3).
    instr(9,return,0).
\end{verbatim}
\end{footnotesize}

The core of the interpreter contains the following clauses.

\begin{footnotesize}
\begin{verbatim}
    interpreter_loop(PC,In,Out) :-
       instr(PC,Opcode,Size), 
       NextPC is PC+Size, 
       format('> ~w  ~w  --> ~w~n',[PC,In,Opcode]),
       ex_opcode(Opcode,NextPC,In,Out).
    ...
    ex_opcode(iconst(Const),NextPC,In,Out) :-
        push(In,Const,Out2),
        interpreter_loop(NextPC,Out2,Out).
    ex_opcode(dup,NextPC,In,Out) :-
        top(In,Top),
        push(In,Top,Out2),
        interpreter_loop(NextPC,Out2,Out).
    ex_opcode(if1(OP,Cst,Label),NextPC,In,Out) :-
        pop(In,RHSVAL1,In2),
        if_then_else(OP,RHSVAL1,Cst,Label,NextPC,In2,Out).
    ex_opcode(iop(OP),NextPC,In,Out) :-
        pop(In,RHSVAL1,In1),
        pop(In1,RHSVAL2,In2),
        ex_op(OP,RHSVAL1,RHSVAL2,Res),
        push(In2,Res,Out2),
        interpreter_loop(NextPC,Out2,Out).
    ex_opcode(return,_,Env,Env).
    ...
       
    if_then_else(OP,Arg1,Arg2,_TrueLabel,FalseLabel,In,Out) :-
        false_op(OP,Arg1,Arg2),
        interpreter_loop(FalseLabel,In,Out).
    if_then_else(OP,Arg1,Arg2,TrueLabel,_FalseLabel,In,Out) :-
        true_op(OP,Arg1,Arg2),
        interpreter_loop(TrueLabel,In,Out).
     
    ex_op(*,A1,A2,R) :- R is A1 * A2.
    ex_op(+,A1,A2,R) :- R is A1 + A2.
    ex_op(-,A1,A2,R) :- R is A1 - A2.
    
    true_op(<=,A1,A2) :- A1 =< A2.
    true_op(>,A1,A2) :- A1 > A2.
    false_op(<=,A1,A2) :- A1 > A2.
    false_op(>,A1,A2) :- A1 =< A2.
    
    pop(env([X|S],Vars),Top,R) :- !, Top=X,R=env(S,Vars).
    pop(E,_,_) :- print('*** Could not pop from stack: '),print(E),nl,fail.
    
    top(env([X|_],_),X).
    
    push(env(S,Vars),X,env([X|S],Vars)).
\end{verbatim}
\end{footnotesize}

We can execute the interpreter for the above bytecode and an initial empty environment (the environment contains
 as first argument the stack and as second argument values for local variables, which we do not use here):
 
\begin{footnotesize}
\begin{verbatim}
| ?- interpreter_loop(0,env([],[]),Out).
> 0  env([],[])  --> iconst(2)
> 1  env([2],[])  --> iconst(2)
> 2  env([2,2],[])  --> iop(*)
> 3  env([4],[])  --> iconst(-1)
> 4  env([-1,4],[])  --> iop(+)
> 5  env([3],[])  --> dup
> 6  env([3,3],[])  --> if1(>,0,3)
> 3  env([3],[])  --> iconst(-1)
> 4  env([-1,3],[])  --> iop(+)
> 5  env([2],[])  --> dup
> 6  env([2,2],[])  --> if1(>,0,3)
> 3  env([2],[])  --> iconst(-1)
> 4  env([-1,2],[])  --> iop(+)
> 5  env([1],[])  --> dup
> 6  env([1,1],[])  --> if1(>,0,3)
> 3  env([1],[])  --> iconst(-1)
> 4  env([-1,1],[])  --> iop(+)
> 5  env([0],[])  --> dup
> 6  env([0,0],[])  --> if1(>,0,3)
> 9  env([0],[])  --> return
Out = env([0],[]) ? 
yes
\end{verbatim}
\end{footnotesize}

As you can see, the above interpreter is {\em deterministic}: when given a state and an opcode it will compute just one solution for
 the successor state after execution of the opcode.
In Prolog one can easily transform such an interpreter into an analysis tool, for either
 data flow analysis \cite{Aho:Dragon2} or abstract interpretation \cite{Cousot92:jlp}.
In that case the Prolog program becomes {\em non-deterministic}.
For example, to transform the above interpreter into an abstract interpreter,
 one has to define abstract operations, such as an abstract multiplication or an abstract ``less or equal'' test, over some abstract domain.
The abstract domain here contains the following abstract values:
\begin{itemize}
 \item {\tt pos} to stand for the positive integers
 \item {\tt neg} to denote the negative integers
 \item {\tt 0} to denote the single value 0
 \item {\tt top} to denote all integers
\end{itemize}
The bottom value is not needed here in the Prolog interpreter; it is represented implicitly by Prolog failure of the interpreter.

\begin{footnotesize}
\begin{verbatim}
    ex_op(*,0,_,0).
    ex_op(*,pos,X,X).
    ex_op(*,neg,0,0).
    ex_op(*,neg,pos,neg).
    ex_op(*,neg,neg,pos).
    ex_op(*,neg,top,top).
    ex_op(*,top,0,0).
    ex_op(*,top,X,top) :- X\=0.
    
    ex_op(+,0,X,X).
    ex_op(+,pos,0,pos).
    ex_op(+,pos,pos,pos).
    ex_op(+,pos,neg,top).
    ex_op(+,pos,top,top).
    ex_op(+,neg,0,neg).
    ex_op(+,neg,pos,top).
    ex_op(+,neg,neg,neg).
    ex_op(+,neg,top,top).
    ex_op(+,top,_,top).
    
    true_op(<=,X,X).
    true_op(<=,top,X) :- X \= top.
    true_op(<=,neg,X) :- X \= neg.
    true_op(<=,0,pos).
    true_op(<=,0,top).
    true_op(<=,pos,top).
    
    true_op(>,_,top).
    true_op(>,_,neg).
    true_op(>,pos,0).
    true_op(>,top,0).
    true_op(>,pos,pos).
    true_op(>,top,pos).
    
    false_op(<=,A1,A2) :- true_op(>,A1,A2).
    false_op(>,A1,A2)  :- true_op(<=,A1,A2).
\end{verbatim}
\end{footnotesize}

\begin{figure}
 \begin{center}
   \includegraphics[width=11.5cm]{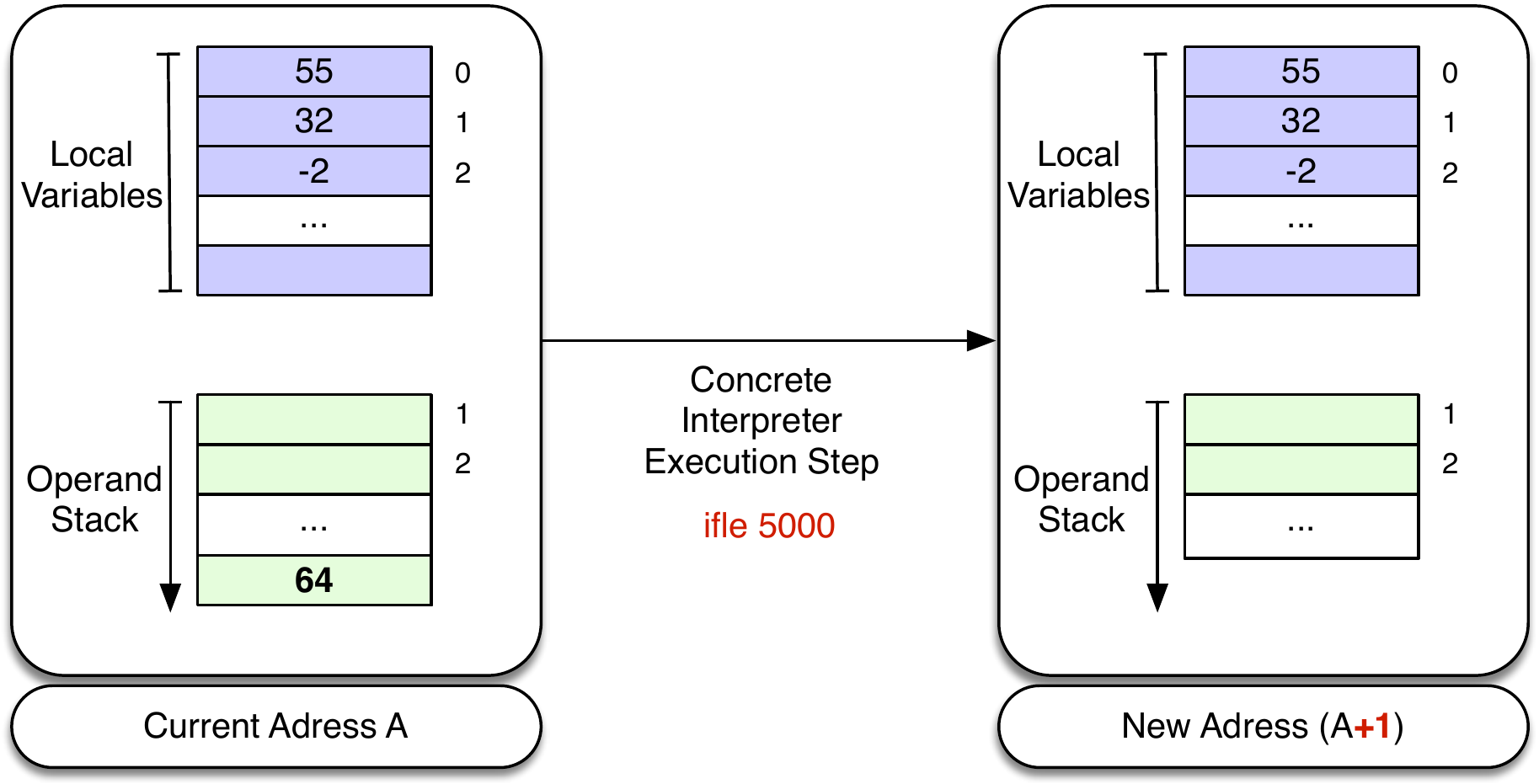} 
   \caption{Deterministic Concrete Interpreter Step
\label{fig:JavaBC-ConcreteStepIf}}
 ~\\
 ~\\
   \includegraphics[width=11.5cm]{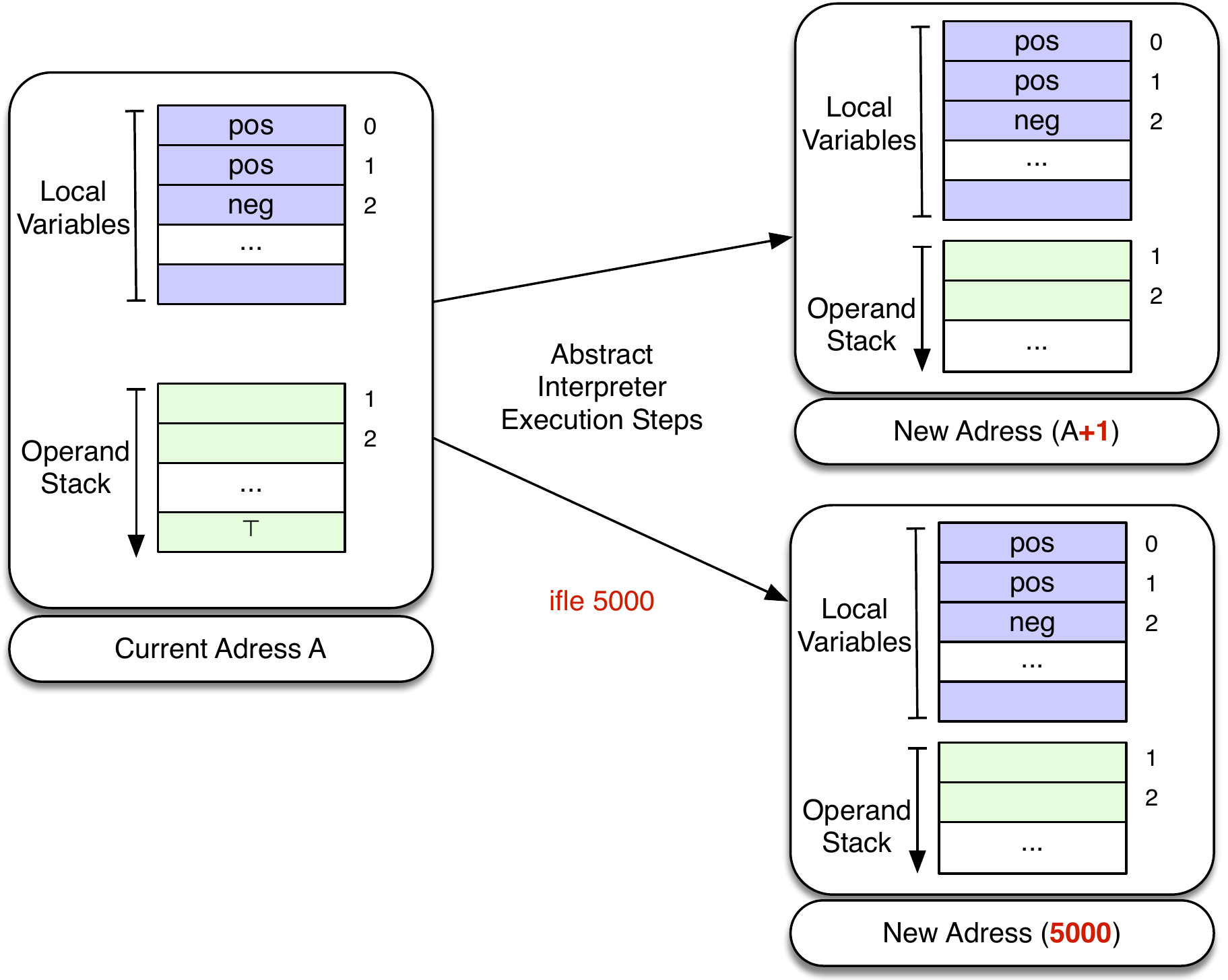}
   \caption{Non-Deterministic Abstract Interpreter Step
\label{fig:JavaBC-AbstractStepIf}}
 \end{center}
\end{figure}

For example, both the call {\tt test\_op(<=,pos,pos)} and {\tt false\_op(<=,pos,pos)} succeeds, and the
 {\tt if\_then\_else} predicate becomes non-deterministic.
This is illustrated in Figures~\ref{fig:JavaBC-ConcreteStepIf} and \ref{fig:JavaBC-AbstractStepIf} for an opcode {\tt ifle 5000},
 which would be encoded {\tt if1(<=,0,5000)} in our Prolog interpreter.
 
To run our interpreter, we first need to use an abstract version of our bytecode program:

\begin{footnotesize}
\begin{verbatim}
    instr(0,iconst(pos),1).
    instr(1,iconst(pos),1).
    instr(2,iop(*),1).
    instr(3,iconst(neg),1).
    instr(4,iop(+),1).
    instr(5,dup,1).
    instr(6,if1('>',0,3),3).
    instr(9,return,0).
\end{verbatim}
\end{footnotesize}

We can now run the same query as above. This time there are infinitely many solutions (paths) through our program.
We show the first three:

\begin{footnotesize}
\begin{verbatim}
 | ?- interpreter_loop(0,env([],[]),R).
> 0  env([],[])  --> iconst(pos)
> 1  env([pos],[])  --> iconst(pos)
> 2  env([pos,pos],[])  --> iop(*)
> 3  env([pos],[])  --> iconst(neg)
> 4  env([neg,pos],[])  --> iop(+)
> 5  env([top],[])  --> dup
> 6  env([top,top],[])  --> if1(>,0,3)
> 9  env([top],[])  --> return
R = env([top],[]) ? ;
> 3  env([top],[])  --> iconst(neg)
> 4  env([neg,top],[])  --> iop(+)
> 5  env([top],[])  --> dup
> 6  env([top,top],[])  --> if1(>,0,3)
> 9  env([top],[])  --> return
R = env([top],[]) ? ;
> 3  env([top],[])  --> iconst(neg)
> 4  env([neg,top],[])  --> iop(+)
> 5  env([top],[])  --> dup
> 6  env([top,top],[])  --> if1(>,0,3)
> 9  env([top],[])  --> return
R = env([top],[]) ? 
\end{verbatim}
\end{footnotesize}

To transform the interpreter into a terminating abstract interpreter one would still need to store
 visited program points and corresponding abstract environments and perform the least upper bound of
 all abstract environments for any given program point (this could be done in the {\tt interpreter\_loop} predicate).

\subsection{Negation} \label{sec-negation}

You may have noticed that in the above concrete interpreter, the {\tt if_then_else} predicate
 was not using Prolog negation {\tt\verb-\+-} or the Prolog if-then-else {\tt ( Tst -> Thn ; Els)}, but used a predicate {\tt true\_op} for a successful comparison
 operator and
 {\tt false\_op} for a failed comparison.
Indeed, the use of the Prolog negation would have prevented the transition from the concrete to the abstract interpreter.

In fact, is rarely a good idea to use Prolog negation to represent negation of the language being analysed.
The reason is that Prolog's built-in negation is not logical negation but so-called ``negation-as-failure''.
This negation can be given a logical description only when its arguments contain no logical variables at the moment
 it is called.
To understand this issue let us examine a simpler program:

\begin{footnotesize}
\begin{verbatim}
    int(0).
    int(s(X)) :- int(X).
\end{verbatim}
\end{footnotesize}

Below are three queries, to this program:

\begin{footnotesize}
\begin{verbatim}
     ?- \+ int(a).   /* succeeds */
     ?- \+ int(X), X=a.  /* fails */
     ?- X=a, \+ int(X).  /* succeeds */
\end{verbatim}
\end{footnotesize}

As you can see in the last two queries, conjunction is not commutative here and the Prolog negation
 is not declarative, i.e., it cannot be described within logic (where conjunction is
  commutative).
More importantly, you can see that the query {\tt int(X)} fails: we cannot use Prolog's negation to
 find values which make a predicate false.
This is what we required in the abstract interpreter above: find values which lead to a comparison operator
 to fail and lead to alternate paths through the bytecode program.

To safely use negation (inside an interpreter) there are basically four solutions.
The first is to always ensure that there are no variables when we call the Prolog negation.
This may be difficult to achieve in some circumstances; and generally means we can use the
interpreter in only one specific way.
For our abstract interpreter above, this means that we cannot use the interpreter to find
values and computation paths which lead to a comparison operator to fail.
 
The second solution is to
   delay negated goals until they become ground.
This can be achieved using the built-in predicate {\tt when/2} of Prolog.
The call {\tt when(Cond,Call)} waits until the condition {\tt Cond} becomes true, at which point {\tt Call} is
 executed.
While {\tt Call} can be any Prolog goal, {\tt Cond} can only use:
  {\tt nonvar(X)}, {\tt ground(X)}, {\tt ?=(X,Y)}, as well as combinations thereof combined with conjunction (,)
   and disjunction (;).
 With this built-in we can implement a safe version of negation, which will ensure that the Prolog negation is
  only called when no variables are left inside the negated call:

\begin{footnotesize}
\begin{verbatim}
  safe_not(P) :- when(ground(P), \+(P)).
\end{verbatim}
\end{footnotesize}

A disadvantage of this approach are refutations which lead to a so-called floundering goal, where all goals suspend.
In that case, one does not know whether the query is a logical consequence of the program or not.
The G\"odel programming language \cite{HillLloyd:Goedel} supported such a safe version of negation.
For program analysis tools, however, we again have the problem that we cannot use this
 kind of negation to find values for variables.

A third solution is to move to another negation, e.g.,
 constructive negation,
 or well-founded or stable model semantics and the associated negation.
This is available, e.g., in answer-set programming \cite{DBLP:books/sp/99/MarekT99} but not in the mainstream Prolog systems. 

Finally, the best solution is to circumvent this problem all together, and use no negation at all.
Here we do this by explicitly writing a predicate for negated formulas.
This is what we have done above, in the form of the predicates {\tt true\_op} and {\tt false\_op}.
Below is an illustration of this approach for a small interpreter for propositional logic:
 
\begin{footnotesize}
\begin{verbatim}
    int(const(true)).
    int(and(X,Y)) :- int(X), int(Y).
    int(or(X,Y)) :- int(X) ; int(Y).
    int(not(X)) :- neg_int(X).
    
    neg_int(const(false)).
    neg_int(and(X,Y)) :- neg_int(X) ; neg_int(Y).
    neg_int(or(X,Y)) :- neg_int(X),neg_int(Y).
    neg_int(not(X)) :- int(X).
\end{verbatim}
\end{footnotesize}

This interpreter now works as expected for negation and partially instantiated queries:

\begin{footnotesize}
\begin{verbatim}
    | ?- int(not(const(X))).
    X = false ? ;
    no
\end{verbatim}
\end{footnotesize}

This interpreter thus actively searches for solutions to the negated formulas.
This technique is used within the \prob{} system to handle negation in the B specification language.
Observe, that had we used Prolog's negation to define the negation in our object programs as 

\begin{footnotesize}
\begin{verbatim}
    int(not(X)) :- \+ int(X).
\end{verbatim}
\end{footnotesize}

\noindent
the answer to the above query would have been {\tt no}.

\subsection{Unification}
\label{sec-unif}
Unification is often useful for looking up information in a program database and to model semantics rules.
For example, in our Java bytecode interpreter above, we can look for conditional jumps to a certain position by unifying with the instruction database:

\begin{footnotesize}
\begin{verbatim}
    | ?- instr(FromPC,if1(_,_,ToPC),_).
    FromPC = 6,
    ToPC = 3 ? 
    yes
\end{verbatim}
\end{footnotesize}

Sometimes unification is not just useful but essential, e.g., when implementing type inference.
Here is a small demo of Hindley-Milner style \cite{Milner:Type78}, written in Prolog with DCGs (Definite Clause Grammars \cite{PereiraWarren:AI80}).
DCGs were initially developed for parsing but are also useful for threading environments in interpreters and in this case type checkers.
Note that these two DCG clauses

\begin{footnotesize}
\begin{verbatim}
    t(a) --> [].
    t(b(A,B)) --> t(A),t(B).
\end{verbatim}
\end{footnotesize}
\noindent
denote this Prolog fact and rule respectively:

\begin{footnotesize}
\begin{verbatim}
    t(a,E,E).
    t(b(A,B),In,Out) :- t(A,In,Env),t(B,Env,Out).
\end{verbatim}
\end{footnotesize}
 
We now encode type inference for a small language containing
 operations ({\tt union}, {\tt intersect}) and predicates({\tt in\_set}) on sets ,
 arithmetic operations ({\tt plus}) and predicates ({\tt gt}), as well as logical conjunction ({\tt and})
 and generic equality ({\tt eq}).

The operators are polymorphic. For example, {\tt eq} can be applied to integers and sets of values.
Similarly, {\tt union} can be applied to sets of values, but in any given set all values must have the same type.
The predicate {\tt type(V,Type,In,Out} holds if the value {\tt V} has type {\tt Type} given the initial
 type environment {\tt In}. The output environment may contain additional variables which are henceforth defined.

\begin{footnotesize}
\begin{verbatim}
    type([],set(_)) --> !, [].
    type(union(A,B),set(R)) --> !,type(A,set(R)), type(B,set(R)).
    type(intersect(A,B),set(R)) --> !,type(A,set(R)), type(B,set(R)).
    type(plus(A,B),integer) --> !,type(A,integer), type(B,integer).
    type(in_set(A,B),predicate) --> !,type(A,TA), type(B,set(TA)).
    type(gt(A,B),predicate) --> !,type(A,integer), type(B,integer).
    type(and(A,B),predicate) --> !,type(A,predicate),type(B,predicate).
    type(eq(A,B),predicate) --> !,type(A,TA),type(B,TA).
    type(Nr,integer) --> {number(Nr)},!.
    type([H|T],set(TH)) --> !,type(H,TH), type(T,set(TH)).
    type(ID,TID) --> {identifier(ID)},\+ defined(id(ID,_)),!,
                      add((id(ID,TID))). % creates fresh variable
    type(ID,TID) --> {identifier(ID)},defined(id(ID,TID)),!.
    type(Expr,T,Env,_) :-
                  format('Type error for ~w (expected: ~w, Env: ~w)~n',[Expr,T,Env]),fail.
    
    defined(X,Env,Env) :- member(X,Env).
    add(X,Env,[X|Env]).
    
    identifier(ID) :- atom(ID), ID \= [].
    
    type(Expr,Result) :- type(Expr,Result,[],Env), format('Typing env: ~w~n',[Env]).
\end{verbatim}
\end{footnotesize}

Note that the {\tt identifier} predicate uses Prolog's negation implicitly in the form of the {\tt\verb+\=+} operator.
This is, however, not an issue as the arguments are all ground.
 
Observe how, e.g., the rule for the union operator requires that the result and
  each argument ({\tt A} and {\tt B})
 is of a set type, but it uses unification of the shared variable {\tt R} to ensure
 that all elements in all sets have the same type (namely {\tt R}).

We can use this small program to perform type inference on the following formula
$$\{z\} \cup \{x,y\} = y \wedge z>v$$

We can correctly determine the types of all
  variables in a single pass:
  
\begin{footnotesize}
\begin{verbatim}
    | ?- type(and(eq(union([z],[x,y]),u),gt(z,v)),R).
    Typing env: [id(v,integer),id(u,set(integer)),id(y,integer),id(x,integer),id(z,integer)]
    R = predicate ? 
    yes
\end{verbatim}
\end{footnotesize}

In some cases, the type inference algorithm will not return a ground type.
E.g., here we compute the type of all variables in the formula $x = \emptyset \cup \emptyset$:

\begin{footnotesize}
\begin{verbatim}
    | ?- type(eq(x,union([],[])),R).
    Typing env: [id(x,set(_1631))]
    R = predicate ? 
    yes
\end{verbatim}
\end{footnotesize}

We see that $x$ is a set, but we do not know what type its elements are.
The program can also be used to generate type error messages:

\begin{footnotesize}
\begin{verbatim}
    | ?- type(and(eq(x,1),eq([],x)),R).
    Type error for x (expected: set(_2167), Env: [id(x,integer)])
    no
\end{verbatim}
\end{footnotesize}

A more complex version of this interpreter is used successfully for type inference for the B specification language within the \prob{} validation tool.

\subsection{Impure Features of Prolog}\label{pure-full-prolog}

While the logical foundations of Prolog --- Horn clauses --- are very elegant
the full Prolog languages contains ``darker'' areas and features which can
only be understood and given meaning when taking the operational semantics
of Prolog into account.
If you look closely at the example in Section~\ref{sec-unif} above, we used the cut (written as {\tt !}), combined
 with a catch-all error clause at the end which always matches,
 to be able to detect typing errors.
The cut here is used to prevent generating type error messages upon backtracking (as the catch-all clause on its own would always be applicable).

In \cite{Le08PPDP} I wrote:
\begin{quote}
 ``Sometimes it is good to view
 Prolog as a dynamic language, and not feel guilty about using the non-ground representation
  or dynamically asserting or retracting facts.
 In many circumstances taking these shortcuts will lead in much shorter and faster code, and it is not clear whether the effort
  in attempting to write a declarative version would be worthwhile.''
\end{quote}

When writing verification or analysis tools in Prolog, it is often a good idea to have
a declarative core, where all predicates $p$ of arity $n$ satisfy for all terms
 $a_1,\ldots,a_n$ and bindings $\theta$:
\begin{itemize}
 \item  $\forall \theta$. $p(a_1,...,a_n), \theta$   $\equiv$  $\theta, p(a_1,...,a_n)$
  (binding-insensitive)
 \item  $p(a_1,...,a_n), \mathit{fail}$   $\equiv$  $\mathit{fail}, p(a_1,...,a_n)$
  (side-effect free)
\end{itemize}
Here $G \equiv H$ means that $G$ and $H$ have the same meaning. Usually, this means the
 same sets of computed answers us

For the infrastructure code (e.g., command-line interface, input-output), it is fine or even mandatory
 to use impure features of Prolog.
For the core of a tool it is also sometimes advisable to use impure features, albeit in a limited fashion.
We should strive to keep the predicates declarative in the absence of error messages.
E.g., as shown in the type-inference program, we can use the non-declarative cut combined with a catch-all to
generate error messages, but the cut did not affect regular type inference.

Co-routines are a mechanism to influence the selection rule of Prolog: via {\tt when} or {\tt block} annotations one can suspend predicate calls until
 a certain condition is met.%
\footnote{A suspended goal is a co-routine; the concept for logic programming dates back to MU Prolog \cite{naish:82}.}
Co-routines can often help to make predicates more declarative, ensuring that they can be used in
multiple directions
Sometimes it is even possible to use non-declarative features to write a declarative predicate.
This is not really surprising, the declarative predicates of the finite domain constraint logic programming library CLP(FD) \cite{CarlssonOttosson:PLILP97} of
 SICStus Prolog \cite{sicstusmanual} are partially implemented in the low-level C-language.
 
Below is the implementation of a declarative addition predicate. 
We use both co-routines and non-declarative features to ensure that the predicate can be used in multiple directions,
is commutative (but still less powerful than addition in CLP(FD)).

\begin{footnotesize}
\begin{verbatim}
    :- block plus(-,?,-), plus(?,-,-), plus(-,-,?).
    plus(X,Y,R) :- ( var(X) -> X is R-Y
                     ; var(Y) -> Y is R-X
                     ; otherwise -> R is X+Y
                    ).
\end{verbatim}
\end{footnotesize}

We have that {\tt plus(1,1,X)} yields {\tt X=2}, while, e.g., {\tt plus(X,1,4)} yields {\tt X=3} as solution.
The block declarations ensure that the predicate {\tt plus} delays until at least two of its arguments are known.
We can thus even solve the equations $x+y=z \wedge z+1=x \wedge x+10=20$.

\begin{footnotesize}
\begin{verbatim}
     | ?- plus(X,Y,Z), plus(Z,1,X), plus(X,10,20).
    X = 10,
    Y = -1,
    Z = 9 ? ;
    no
\end{verbatim}
\end{footnotesize}

The first two calls to {\tt plus} will initially be suspended, while the third call {\tt plus(X,10,20)} will
 be executed, instantiating {\tt X} to 10. This will unblock the second call {\tt plus(Z,1,X)},
  instantiating {\tt Z} to 9, which in turn will unblock the first call to {\tt plus}.
  

\section{Conclusion: An Assessment of Key Prolog Technologies for Verification and Analysis Tools} \label{sec-assessment}

Non-determinism and unification of Prolog is useful and as shown sometimes essential, e.g., for type inference.
I find {\bf co-routines} (when and block) to be
absolutely essential for Prolog applications in : custom constraint solver, writing declarative reversible predicates, 
 or implementing the Andorra principle (see also \cite{Le08PPDP}).

Similarly, {\bf constraint logic programming} (CLP) is an important feature of modern Prolog systems for many applications.
I found the finite domain library CLP(FD) \cite{CarlssonOttosson:PLILP97} to be the most useful.
I have not really used  the boolean constraint solver CLP(B) of SICStus Prolog.
For my particular use cases within \prob{} its encoding using binary decision diagrams (BDDs \cite{Bryant:ACMSurv92}) was too slow
and I resorted to writing my own boolean solver using attributed variables.

{\bf Attributed variables} allow one to attach attributes to logical variables.
One then provides hooks which are called by Prolog when variables with attributes are unified.
Attributed variables are a good fallback solution for writing custom constraint solvers, but they are very low-level and hence tricky to write and debug.

Constraint Handling Rules ({\bf CHR}) \cite{Fruewirth:CHRbook} can
take the pain out of dealing with attributed variables.
CHR provides a higher-level way of writing constraint solving rules.
However, I also found CHR code quite
 difficult to debug and found it difficult to write larger solvers which perform well (and do not loop).
Currently CHR is used optionally in \prob{} for some integer arithmetic propagation rules, but it is not used heavily.

{\bf Tabling} \cite{ChenWarren:JACM96} can be very useful in the context of verification and program analysis.
It is, however, tricky in the context of constraints and co-routines and is not provided, e.g., by SICStus Prolog.
Tabling was used in the XTL \cite{LeuschelMassartCurrie:FME2001} and XMC \cite{RamakrishnaEtAl:CAV97,XMC:CAV2000} model checkers.
In my case, the combination of tabling and obtaining computed answers inside negation (cf. Section~\ref{sec-negation})
 was tricky to achieve and prevented using this approach
 for more complex specification language.
Within the \prob{} tool, tabling is implemented in an ad-hoc manner in various instances;
  in the end the comfort of SICStus and its CLP(FD) library and co-routines turned out to be more important than
  efficient tabling.
But I would still wish to have access to a Prolog system which combines all of these features.

\paragraph{Prolog: The Missing Bits}

The absence of a full-fledged Prolog type checker is an issue.
A solution is to make systematic use of catch-all clauses, as shown in Section~\ref{sec-unif}.
Generally, this should be complemented with extensive 
 unit tests, runtime tests, and integration tests with continuous integration.
For the latter a command-line interface is very useful.
I also found the implementation of a REPL (Read-Eval-Print-Loop)%
\footnote{See, e.g., {\tt \url{https://en.wikipedia.org/wiki/Read-eval-print_loop}}.}
 for the object language to be a useful
 way to generate new unit tests.
When using co-routines, it can be useful to programmatically 
 generate unit tests, varying the order in which arguments are instantiated from the outside.

There are still a few aspects of verification and analysis which are difficult to implement efficiently
in Prolog.
One is loop checking in a graph:
as already discussed in \cite{Le08PPDP}, the
 LTL model checker of \prob{} written in C for this reason.

Similarly,
  SICStus Prolog unfortunately does not yet provide built-in hash-maps or similar data structures.
This was relevant for implementing directed model checking, where one uses a heuristic function to prioritise the unprocessed states which should be checked next.
Within \prob{} \cite{DBLP:conf/sbmf/LeuschelB10} we resorted to (partially) storing this 
 queue of unprocessed states in a C++ multimap, which enabled us to quickly add new states and obtain the state with highest priority. 
 
The support for parallel execution in Prolog is rather patchy and often limited.
For implementing parallel and distributed model checking in \prob{} \cite{DBLP:conf/nfm/KornerB18},
 multiple Prolog instances were used, communicating via
 ZMQ \cite{0mq} provided to SICStus Prolog via its C interface. 

Finally,
 compared to languages like Java or JavaScript, Prolog systems only have access to relatively few standard libraries.
This meant for example that we used the C++ regular expression library in \prob{} to provide a regular expression
 library for B specifications.

 In summary, despite its shortcomings, Prolog is still an excellent language to implement verification and
  program analysis and transformation techniques and tools.
In particular when it comes to traversal and manipulation of abstract syntax trees, Prolog programs
 are in my experience more compact, faster and more memory efficient than many programs written
 in more mainstream languages like Java.

\subsection*{Acknowledgements}
I would like to thank Laurent Fribourg for his useful feedback on an earlier version of the article.

\bibliographystyle{eptcs}
\bibliography{michael}
\end{document}